\documentclass[epj]{svjour}
\usepackage{graphics}
\usepackage{amsmath,amssymb}

\begin{document}

\title{Fractional Bhatnagar-Gross-Krook kinetic equation}

\author{Igor Goychuk \thanks{\emph{E-mail:} igoychuk@uni-potsdam.de}}

\institute{Institute of Physics and Astronomy, University of Potsdam,
Karl-Liebknecht-Str. 24/25, 14476 Potsdam-Golm, Germany}
\date{Received: date / Revised version: date}

\abstract{The linear Boltzmann equation approach is generalized to describe fractional
superdiffusive transport of the L\'{e}vy walk type in external force fields.  The
time distribution between scattering events is assumed to have a finite mean value and
infinite variance. It is completely characterized  by the two
scattering  rates, one fractional and a normal one, which defines
also the mean scattering rate. We formulate a
general fractional linear Boltzmann equation approach and exemplify it with a
particularly simple case of the Bohm and Gross scattering integral leading 
to a fractional generalization of the Bhatnagar, Gross and Krook kinetic equation. 
Here, at
each scattering event the particle velocity is completely randomized and takes a
value from equilibrium Maxwell distribution at a given fixed temperature.  We
show that the retardation effects are indispensable even in the limit of infinite mean
scattering rate and
argue that this novel fractional  kinetic equation provides a viable alternative
to the fractional Kramers-Fokker-Planck (KFP) equation by Barkai and Silbey and its  
generalization 
by Friedrich \textit{et al.} based on the picture of
divergent mean time between scattering events. The case of divergent mean
time is also discussed at length and compared with the earlier results
obtained within the fractional KFP.
}
\authorrunning{Igor Goychuk}
\titlerunning{Fractional Bhatnagar-Gross-Krook kinetic equation}
\maketitle

\section{Introduction}

Linear Boltzmann equation \cite{Hoare} provides 
a versatile tool to describe kinetics of the test or impurity
particles in a background gas of abundant particles serving as a thermal bath. 
Consider test particles of mass $M$ characterized by
the distribution function $f(x,v,t)$, where $x$ is the particle position, $v=\dot x$ is its velocity
and $t$ is time. The particles are subjected to an external force $F(x,v,t)$ and 
obey the Newtonian equations of motion $M\ddot x=F(x,v,t)$, where $x$, $v$, and $F$ are 
generally vectors with
three Cartesian components. In this paper, we are dealing
for simplicity with a one-dimensional scalar form. However, the major results can easily
be generalized to higher dimensions. 
From time to time the impurity particles are scattering with
the background gas particles and their velocities are changed at each scattering event.
Any kinetic equation can be written in the form 
$df(x,v,t)/dt={\rm St}[f(x,v,t)]$, where $d/dt$ is a full derivative and ${\rm St}[f(x,v,t)]$
is a scattering integral (Sto\ss integral), or collision term. In the absence of scattering events,
$df(x,v,t)/dt=0$ expresses just the Liouville theorem of classical mechanics, or conservation
of the number of particles in an elementary phase volume $dxdp$, $p=Mv$. This number
changes due to scattering events yielding income and outcome of the particles in any given
phase volume. At the equilibrium,
the both processes are balanced, and hence ${\rm St}[f_{\rm eq}(x,v)]=0$, where 
$f_{\rm eq}(x,v)$ is equilibrium distribution. Generally, ${\rm St}[f(x,v,t)]$ is a 
nonlinear function of $f(x,v,t)$, like in the classical nonlinear Boltzmann equation, which
takes binary collisions between the particles of one-component gas into account. However,
in the case of a two-component gas, when the gas of the test particles is very dilute, 
the scattering events among the test particles
can be simply neglected because they are very rare, 
and this leads to a linear Boltzmann equation (LBE),
\begin{eqnarray}\label{LBE}
\frac{\partial f(x,v,t)}{\partial t}&&+
v\frac{\partial f(x,v,t)}{\partial x}+\frac{F(x,v,t)}{M}
\frac{\partial f(x,v,t)}{\partial v}\\
=&&-r(v)f(x,v,t)+\int dv'w(v\leftarrow v')f(x,v',t), \nonumber
\end{eqnarray}
where $w(v'\leftarrow v)$ is the rate of transitions in the velocity space, and
$r(v)=\int dv' w(v'\leftarrow v)$ is a total rate.
The first line presents the full derivative of the distribution function, and
the second line is a particular linear form of the scattering integral. 
It says that the scattering in the velocity subspace is described by
a standard master equation.
In writing this equation one implicitly assumes that the time between the 
scattering events is exponentially distributed. This is the reason why the
scattering integral is local in time, and underlying dynamics is Markovian. 
The standard Kramers-Fokker-Planck equation presents a limiting case of LBE, where a diffusional
approximation is done in the scattering integral \cite{Hoare,Risken}, 
which we write in the following generic form 
consistent with thermodynamics \cite{Klimontovich}
\begin{eqnarray}\label{KK}
{\rm St}[f(x,v,t)]&=&\frac{\partial}{\partial v}\left ( D(v) e^{-\beta Mv^2/2}
\frac{\partial}{\partial v}  e^{\beta Mv^2/2}f(x,v,t)\right )\nonumber \\
&=&\frac{\partial}{\partial v}\left [ \gamma(v) \left( v+v_T^2
\frac{\partial}{\partial v}  \right) f(x,v,t)\right ],
\end{eqnarray}
where $\beta=1/(k_BT)$ is inverse temperature, $D(v)=v_T^2\gamma(v)$, 
$\eta(v)=M\gamma(v)$ is a nonlinear, generally velocity-dependent friction coefficient, and 
$v_T=\sqrt{k_BT/M}$ is thermal velocity.
This form emerges with the help of the Kramers-Moyal expansion \cite{vanKampen} 
in the master equation. For example, in the particular case
of $\gamma(v)=\gamma=const$ one can use a modified Rayleigh model of the 
scattering kernel  
\begin{eqnarray}\label{form}
w(v\leftarrow v')=\frac{1}{4}\gamma \kappa\sqrt{\frac{\kappa}{2\pi v_T^2}}
 e^{-\frac{\kappa(v-v'+(v'+v)/\kappa)^2}{8v_T^2}}\;,
\end{eqnarray}
where $\kappa=M/m$ is the ratio of the masses of the test and background particles.
This is a modification of the standard Rayleigh kernel, see Eq. (3.4) in \cite{Hoare}
or Eq. (4.14) in \cite{vanKampen}, 
where differently from the standard model we assume that the frequency of collisions
does not depend on the relative velocity of the test particles and the particles
of the thermal bath. For this reason, the prefactor in (\ref{form}) does not contain
the difference of velocities $|v-v'|$.
In this modified Rayleigh model,
the total collision rate is $r(v)=r=(1/2)\gamma\kappa^2/(\kappa+1)$.
Eq. (\ref{KK}) is obtained \textit{exactly} with $\gamma(v)=\gamma=const$ from the scattering
integral in Eq. (\ref{LBE}) upon using the Kramers-Moyal expansion and taking
the limit $\kappa\to\infty$ 
\footnote{With $\bar w(v';u):=w(v\leftarrow v'), u=v-v'$, the first
two Kramers-Moyal coefficients, $a_{j}(v)=\int_{-\infty}^{\infty}u^j\bar w(v;u)du$ \cite{vanKampen},
read: $a_1(v)=-\gamma v\kappa^2/(1+\kappa)^2$, 
$a_2(v)=2\gamma v_T^2\kappa^3/(1+\kappa)^3+2\gamma v^2\kappa^2/(1+\kappa)^3$,
and all $a_{j>2}=o(1/\kappa)$ vanish in the $\kappa\to\infty$ limit.}. 
Notice that then also $r\to\infty$.
This corresponds physically to the case where the scattering events occur very often,
and the background gas of light particles is dense and fluid-like (heavy 
Brownian particles in a fluid).
The first line in (\ref{KK}) makes it immediately clear that this equation is compatible
with the thermal equilibrium, where $f_{\rm eq}(x,v)=p_x(x)f_M(v)$ and
\[
f_M(v)=\exp[-v^2/(2v_T^2)]/\sqrt{2\pi v_T^2}
\] 
is the equilibrium Maxwellian velocity distribution.

Another important instance of the LBE equation is provided by a Bohm and Gross form of the
scattering integral \cite{Bohm}. It can be obtained from the modified Rayleigh model (\ref{form})
in the case $\kappa=1$, i.e. the test particles and the particles of the thermal bath
have equal masses.
In this case,
 \[w(v\leftarrow v')=rf_M(v), \] 
with the collision rate $r=\gamma/4$.
The physical meaning of this choice is as follows. 
Time-intervals between
scattering events are exponentially distributed with the mean time
$\langle \tau\rangle=1/r$, and after each scattering event the particle's velocity
is fully randomized in the correspondence with its thermally equilibrium distribution $f_M(v)$.
The scattering integral in this case reads \cite{BGR},
\begin{eqnarray}\label{BG}
{\rm St}[f(x,v,t)]=-r \left [f(x,v,t)-f_M(v)\int_{}^{} dv f(x,v,t) \right],
\end{eqnarray}
and the corresponding kinetic equation is known as Bhatnagar, Gross and Krook (BGK)
kinetic equation \cite{BGR,Risken,Zwanzig}. This one is considered typically
as a linear approximation to a nonlinear Boltzmann equation, where the distinct 
background and impurity particles have yet nearly equal masses.
In  the kinetic equation for the reduced distribution function of velocities,
 $p(v,t)=\int dx f(x,v,t)$ in the velocity subspace,
 the scattering term looks especially
simple,  $-r \left [p(v,t)-f_M(v)\right]$, which corresponds
to a single relaxation time approximation ($F=0$ here), 
\[
\partial p(v,t)/\partial t=
-r \left [p(v,t)-f_M(v)\right]. \]
In particular, because of this simplicity,
  BGK kinetic equation became popular in the literature \cite{Zwanzig}, especially in the
  context of lattice Boltzmann models \cite{Succi} 
  aimed at the lattice simulations of hydrodynamics. 
  In this respect, derivation of the hydrodynamics
  equations from the BGK kinetic equation is especially simple and insightful \cite{Zwanzig},
  what underlines its general importance and a possibly wide range of applications
  beyond gaseous systems like plasmas.

It is the main purpose of this paper to generalize this linear Boltzmann equation
description towards a fractional L\'{e}vy walk kinetics 
in the velocity space
\cite{Kenkre73,Shlesinger,Scher,Bouchaud,Hughes,Balescu,BenAvraham,Geisel88,Zumofen93,ZumofenPhysicaD,West97,Barkai96,Barkai97,BarkaiPRE97,Metzler00}, 
where the times between scattering events
are 
non-exponentially distributed and possess a finite first moment, i.e. the mean
time $\langle \tau \rangle $ between the scattering events remains finite 
\cite{Geisel88,Zumofen93,ZumofenPhysicaD,West97,Barkai97,BarkaiPRE97,GoychukPRE12}. Moreover,
we will pay an essential attention to the limit $\langle \tau \rangle\to 0$.
In this respect, our theory differs much from the fractional Kramers-Fokker-Planck (KFP) equation 
by
Barkai and Silbey \cite{Barkai00}, and its further correction and generalization by 
Friedrich \textit{et al.} \cite{FriedrichPRL06,FriedrichPRE06} based on the picture of infinite 
$\langle \tau \rangle \to \infty$, even if it is related closely
in several aspects to the theory developed in \cite{FriedrichPRL06,FriedrichPRE06}.  

\section{Theory}

We start from considering the scattering process as a 
continuous time random walk (CTRW) \cite{Kenkre73,Shlesinger,Scher,Bouchaud,Hughes,Balescu,BenAvraham} 
in the velocity space or as a 
L\`{e}vy walk 
\cite{Geisel88,Bouchaud,Zumofen93,ZumofenPhysicaD,Hughes,Balescu,BenAvraham,Barkai96,West97,Barkai97,BarkaiPRE97,Metzler00}.
The particles fly with a constant velocity $v$ between any two subsequent 
scattering events and such events change their velocity from $v$ to
$v'$ with a transition probability density $W(v'\leftarrow v)$. We will consider
the case, where the mean time between the scattering events exists and it defines
the mean scattering rate $r=1/\langle \tau \rangle $. Then, 
$w(v'\leftarrow v)=rW(v'\leftarrow v)$. All scattering events are assumed to
be mutually independent.

\subsection{L\'{e}vy walk in the velocity space}

In the velocity space, such a decoupled semi-Markovian L\'{e}vy walk 
is fully characterized by the residence time distribution  or
RTD $\psi(\tau)$ of the time-intervals between two scattering events 
and the transition probability density 
$W(v'\leftarrow v)$. We consider first the dynamics of the velocity
distribution $p(v,t)=\int dx f(x,v,t)$. It is governed
by a generalized master equation (GME), which is well-known
by analogy with such a decoupled CTRW in the coordinate space.
This GME reads \cite{Kenkre73} 
\begin{eqnarray}\label{GME}
&&\frac{\partial p(v,t)}{\partial t}= -r(v)\int_0^{t} dt'K(t-t')p(v,t') \\
&& +\int_0^{t}  dt'K(t-t')\int dv'w(v\leftarrow v')p(v',t')\;,\nonumber
\end{eqnarray}
with a memory kernel $K(t)$ whose Laplace-transform is
$\tilde K(s)=(1/r)s\tilde \psi(s)/[1-\tilde \psi(s)]=(1/r)\tilde \psi(s)/\tilde \Phi(s)$,
in terms of the Laplace-transformed RTD $\tilde \psi(s)$. It can be expressed also
through the survival probability $\Phi(\tau)=\int_{\tau}^{\infty}\psi(t)dt$, which is
the probability to do not have any scattering event within a time interval of 
length $\tau$. If this survival probability is exponential, $\Phi(\tau)=\exp(-rt)$,
then $K(t)=\delta(t)$ and Eq. (\ref{GME}) is the standard LBE for the force-free case
in the velocity space.

We consider a generalization of this LBE, where the RTD between two scattering
presents a sum, \\ $\psi(\tau)=\sum_{j=0}^N\psi_{j}(\tau)$
over $N+1$ independent scattering channels, and the corresponding
Laplace-transforms read \cite{GoychukPRE12}
\begin{eqnarray}\label{2}
\tilde \psi_{j}(s)=r_{\alpha_j} s^{1-\alpha_j}\tilde \Phi(s),\;
\end{eqnarray} 
where 
\begin{eqnarray}\label{1}
\tilde \Phi(s)=\frac{1}{s+\sum_{j=0}^N r_{\alpha_j} s^{1-\alpha_j}}\;
\end{eqnarray} 
with $0< \alpha_j\leq 1$ is the Laplace-transformed survival probability  
$\Phi(t)$, and $ r_{\alpha_j} $ are
the fractional scattering rates. We demand that one of them is normal, $\alpha_0=1$,
and at least one of them is anomalous.
The normal rate $ r_{\alpha_0}=1/\langle \tau\rangle=r$ defines the mean scattering rate,  
The variance of $\tau$ in this model is infinite due
to anomalous scattering channels. 
This distribution has been derived in Ref. \cite{GoychukPRE12}
in assumption that each of the scattering channels taken separately is characterized by
a Mittag-Leffler distribution 
$\psi_j^{(sep)}(\tau)=-(d/d\tau)E_{\alpha_j}(-r_{\alpha_j}\tau^{\alpha_j})$,
where $E_\alpha(z)=\sum_{n=0}^\infty z^n/\Gamma(n\alpha+1)$ is the Mittag-Leffler function,
and one of the independent channels is taken randomly at each scattering event, i.e. they are
acting intermittently and in parallel. Here, 
$\Gamma(x)$ is a standard special $\Gamma$-function. 
The averaged number of the
scattering events in this model grows as 
$\langle n(t)\rangle =\sum_{j}r_{\alpha_j}t^{\alpha_j}/\Gamma(1+\alpha_j)$.
For simplicity, we will
restrict our attention to the model with one normal and one fractional scattering rates. 
Then, $\langle n(t)\rangle =rt+r_{\alpha}t^\alpha/\Gamma(1+\alpha)$, exactly. Notice,
that with respect to the averaged number of the scattering events, an anomalous scattering
channel contributes really a little for sufficiently large $t$. 
However, its role in the kinetics is really profound!

We wish to find the diffusional spread of the variance of the particles position 
$\langle x^2(t)\rangle $ assuming that at the initial time $t_0=0$ they all were localized
at the coordinate origin, $x(0)=0$. For this, we need to know the velocity autocorrelation
function (ACF) of two arguments $K_v(t,t')=\langle v(t')v(t'')\rangle$. 
Indeed, by using $x(t)=\int_0^t v(t')dt$, we have
$\langle x^2(t) \rangle=\int_0^t dt'\int_0^{t}dt'' \langle v(t')v(t'')\rangle$.
To find such a nonstationary, or aging velocity ACF is not a trivial task 
\cite{Cox,Godreche,Allegrini05,Margolin05,Froemberg13}.
For example,
in the case of two-state velocity fluctuations it was solved in Ref. \cite{Froemberg13}.
A further simplification is possible for the case of a stationary
ACF $K_v^{(\rm st)}(|t-t'|)$, which depends only on the difference of two 
time arguments. Then, 
\begin{eqnarray}\label{diff1}
\langle x^2(t) \rangle&&=2 \int_0^t dt'\int_0^{t'}dt''K_v^{(\rm st)}(|t'-t''|)\\
&&=
2\int_0^t (t-t')K_v^{(\rm st)}(t')dt'.\nonumber
\end{eqnarray}
 In this important case, the Laplace-transform
of   $\langle x^2(t) \rangle$ reads 
$\widetilde{\langle x^2(s) \rangle}=2 \tilde K_v^{(\rm st)}(s)/s^2 $, where
$\tilde K_v^{(\rm st)}(s)$ is the Laplace-transform of $K_v^{(\rm st)}(\tau)$.

It must be stressed, however, that the GME  (\ref{GME}) corresponds to a CTRW, 
which starts at the time $t_0=0$
from a scattering event \cite{Tunaley}. As a matter of fact, if to calculate the ACF
$K_v(t,0)$ for $t\geq 0$ using this master equation and
an initially equilibrium $p(v,0)=f_M(v)$, we obtain not the stationary ACF 
$K_v^{(\rm st)}(t)$, but just  a non-stationary
velocity ACF $K_v(t_{\rm ag}+t,t_{\rm ag})$ taken at zero age time 
$t_{\rm ag}=0$ \cite{Cox}.
This one simply cannot be used in Eq. (\ref{diff1}). To use it therein, would
be a profound mistake. To find $K_v(t_{\rm ag}+t,t_{\rm ag})$, one needs
to consider an aging CTRW, where the survival probability of the first
scattering time interval $\Phi^{(0)}(t|t_{\rm ag})$ is different and age-dependent. 
It can be found from the following reasoning. 
Assume that scattering events started at the time $-t_{\rm ag}$ in the past relative
to the starting point  $t_0=0$ of observations. Then, 
if we observe our system from $t_0=0$ to $t$, the corresponding survival probability
to do not have a scattering event is $\Phi(t+t_{\rm ag})$. 
However, $n$ undetected scattering  events
might already took place until any ``unseen'' 
time point $-y$ within the time interval $[-t_{\rm ag},0]$ 
in the past and then
no events occurred until $t$. Integrating over $y$ and summing over all possible $n$ yields
the following exact result \cite{Cox,Godreche,Allegrini05}
\begin{eqnarray}\label{age1}
\Phi^{(0)}(t|t_{\rm ag})=&&\Phi(t+t_{\rm ag}) \\
+&&\sum_{n=1}^{\infty}\int_0^{t_a} \psi_{n}(t_a-y)\Phi(t+y)dy \nonumber,
\end{eqnarray}
where $\psi_{n}(t)$ is the probability density to have $n$ scattering events. It is the $n$-time
convolution of the density $\psi(\tau)$, $\tilde \psi_n(s)=[\tilde\psi(s)]^n$ in the Laplace space.
From (\ref{age1}), one can easily find the double Laplace transform 
$\tilde  \Phi^{(0)}(s|u)$ of $\Phi^{(0)}(t|t_{\rm ag})$, 
where $s$ is the Laplace-transform variable, which is conjugated to $t$, and $u$ to $t_{\rm ag}$.
Some  algebra yields simple result \cite{GoychukCTP14}
\begin{eqnarray}
\tilde  \Phi^{(0)}(s|u)=\frac{1}{u(s-u)}\left( 1-\frac{\tilde\Phi(s)}{\tilde\Phi(u)}\right)\;.
\end{eqnarray}

The Laplace-transform of the fully aged or equilibrium
$\Phi^{(0)}(t)=\lim_{t_{\rm ag}\to\infty}\Phi^{(0)}(t|t_{\rm ag})$, 
the first-time stationary survival probability, 
can be obtained now as  
$\tilde \Phi^{(0)}(s)=\lim_{u\to 0}u \tilde  \Phi^{(0)}(s|u)$. For 
$\tilde\Phi(0)=\langle \tau\rangle\neq \infty$ this yields the well-known result  
$\tilde \Phi^{(0)}(s)=(1- \tilde \Phi(s)/\langle \tau\rangle)/s$
\cite{Cox,Tunaley,GoychukPRL03,GoychukPRE04,GoychukCTP14} . The corresponding
first-interval RTD is $\psi^{(0)}(t)=\Phi(t)/\langle \tau\rangle$, 
which is also well-known.
Using this $\psi^{(0)}(t)$ one can derive another GME,
which corresponds to a time-homogeneous initial preparations of the scattering
process. This was done in the Appendix of Ref. \cite{GoychukPRE04}, in a different
context. Applying that GME to our scattering process we obtain,
\begin{eqnarray}\label{GME2}
&&\frac{\partial p(v,t)}{\partial t}= 
-r(v)\int_0^{t} dt'K(t-t')[p(v,t')-p(v,0)]\nonumber \\
&& +\int_0^{t}  dt'K(t-t')\int dv'w(v\leftarrow v')[p(v',t')-p(v',0)]\;\nonumber\\
&& -r(v)p(v,0)+\int dv' w(v\leftarrow v') p(v',0)\;.
\end{eqnarray}
This equation corresponds to the following initial preparation. We first trap
the particles in a space trap 
and let them
pre-equilibrate with the particles of the thermal bath by multiple collisions
before we release them from the trap. The distribution of velocities
at $t_0=0$ can still be out of equilibrium. This is what van Kampen named
as ``extraction of a subensemble'' \cite{vanKampen}.
We are doing such a procedure for non-Markovian renewal processes with finite $\langle \tau\rangle$.
Notice that in the limit $\langle \tau\rangle\to \infty$ such an extraction
of subensemble is not possible in principle, because the corresponding random process
simply does not have even a wide-sense stationary limit.
The process $v(t)$ is still non-stationary for $p(v,0)\neq p_{\rm st}(v)$. However, the memory
kernel of transition probabilities depends now merely on the time shift indeed.
Very important is that $K_v(t,0)$ found with the help of 
GME (\ref{GME2}) is indeed the stationary velocity ACF, $K_v(t,0)=K_v^{(\rm st)}(t)$,
$t\geq 0$. Namely, the solution of (\ref{GME2}) with $p(v,0)=\delta(v-v')$
yields the time-shift invariant propagator of velocities $\Pi^{\rm (st)}(v,t|v',0)$, and
\begin{eqnarray}\label{calc}
K_v^{\rm (st)}(t)=\int \int vv'\Pi^{\rm (st)}(v,t|v',0)p_{\rm st}(v')dvdv'\;,
\end{eqnarray}
where $p_{\rm st}(v)=\lim_{t\to\infty}p(v,t)$ is the stationary solution.

Next, our fractional  BGK master equation  
is characterized by the RTD having the characteristic function
or the Laplace transform
\begin{eqnarray}\label{RTD}
\tilde \psi(s)=\frac{r+r_{\alpha} s^{1-\alpha}}{s+r+r_{\alpha} s^{1-\alpha}},\;
\end{eqnarray} 
and by a complete randomization of the velocity at each scattering event in accordance
with the Maxwellian distribution of velocities, like in the kinetic 
model of Refs. \cite{Barkai97,BarkaiPRE97}. Arguably, this is a simplest
fractional generalization of the kinetic BGK model possible. 
Notice, that $\psi(t)$ has $\psi(t)\propto t^{-3+\alpha}$ long-time asymptotics\footnote{
This should by kept in mind while comparing our asymptotic results
with other earlier published results which used another parameterization,
$\psi(t)\propto t^{-1-\gamma}$ with $1<\gamma<2$. Then, our $\alpha=2-\gamma$. } 
within this model \cite{GoychukPRE12,GoychukCTP14} .   The Laplace-transform
of the survival probability reads $\tilde \Phi(s)=1-s\tilde \psi(s)$ or
\begin{eqnarray}\label{surv1}
\tilde \Phi(s)=\frac{1}{s+r+r_{\alpha} s^{1-\alpha}},\;
\end{eqnarray} 
and the first-time survival probability is
\begin{eqnarray}\label{surv2}
\tilde \Phi^{(0)}(s)=\frac{1+r_{\alpha} s^{-\alpha}}{s+r+r_{\alpha} s^{1-\alpha}}\;
\end{eqnarray} 
in the Laplace-domain.
Furthermore, the GME (\ref{GME}) yields a time-inhomogeneous fractional BGK kinetic
equation in the velocity space \cite{GoychukCTP14}
\begin{eqnarray}\label{v-space}
\frac{\partial p(v,t)}{\partial t}=-\left (r+r_{\alpha}\sideset{_0}{_t}{\mathop{\hat
D}^{1-\alpha}}\right )\left [p(v,t)-f_M(v) \right ],
\end{eqnarray}
where $\sideset{_0}{_t}{\mathop{\hat
D}^{1-\alpha_j}}$
is the operator of Riemann-Liouville fractional derivative defined as \cite{Metzler00,Gorenflo}
\begin{eqnarray}\label{RL} 
\sideset{_{0}}{_t}{\mathop{\hat D}^{\gamma}}f(t):=\frac{1}{\Gamma(1-\gamma)}
\frac{\partial}{\partial t}\int_{0}^t dt' \frac{f(t')}{(t-t')^\gamma}
\end{eqnarray}
by its action on a test function $f(t)$, $0<\gamma<1$. 
This is an example of kinetic equations with fractional
derivatives of distributed order \cite{distro}. 
Moreover, the GME (\ref{GME2}) for this model yields
\begin{eqnarray}\label{v-space2}
\frac{\partial p(v,t)}{\partial t}=-r_{\alpha}\sideset{_0}{_t}{\mathop{\hat
D}^{1-\alpha}}\left [p(v,t)-p(v,0) \right ]\\
-r\left [p(v,t)-f_M(v) \right ]\;.\nonumber
\end{eqnarray}

\section{Fractional superdiffusion from the velocity space perspective}

We proceed further with showing that the considered description
in the velocity space does yield fractional superdiffusion
in the  coordinate space. For this, we calculate 
$K^{\rm (st)}_v(|t-t'|)$. Here, we can use
the solution of Eq. (\ref{v-space2}), which is easy to obtain in the Laplace-domain.
In the time-domain, it reads
\begin{eqnarray}\label{solution}
p(v,t)=f_M(v)+\Phi^{(0)}(t)[p(v,0)-f_M(v)]\;,
\end{eqnarray}
like in \cite{Barkai97}.
Hence, the stationary propagator or \textit{stationary, time-shift invariant} 
conditional probability of the velocity
distribution reads $\Pi^{\rm (st)}(v,t|v',0)=f_M(v)+
\Phi^{(0)}(t)[\delta(v-v')-f_M(v)]$, and with $p_{\rm st}(v)=f_M(v)$ in (\ref{calc})
we obtain
\begin{eqnarray}\label{Kv_st}
K_v^{\rm (st)}(|t-t'|)=v_T^2\Phi^{(0)}(|t-t'|)\;.
\end{eqnarray}
Notice a remarkable simplicity if this result within the studied scattering model:
the normalized stationary velocity autocorrelation function just equals to the
equilibrium survival probability of the first scattering time intervals. A similar
result was obtain within a mathematically related model for dielectric relaxation
which describes a stationary generalized Cole-Cole response \cite{GoychukCTP14}, and also
earlier within a different model with just two-state, $\pm v$, 
velocity fluctuations \cite{Geisel88,West97}.
By the same token, we find upon the use of Eq. (\ref{v-space}), 
$K_v(t,0)=v_T^2\Phi(t)$. 
Notice once again that this  later $K_v(t,0)$
cannot be used to find $\langle x^2(t)\rangle$.
This is a nonstationary velocity ACF for zero aging time.

\subsection{The limit $\langle \tau \rangle\to\infty$ or $r\to 0$}

In this limit, which alludes to the fractional KFP equation of 
Refs. \cite{Barkai00,FriedrichPRL06,FriedrichPRE06}, we obtain from Eq. 
(\ref{surv2}), $\Phi^{(0)}(t)=1$, like in Ref. \cite{GoychukCTP14}. 
Hence, $K_v^{\rm (st)}(|t-t'|)=v_T^2$, i.e.
it does not decay at all!
In accordance with the Slutsky-Khinchine theorem \cite{Papoulis}, 
this means that such a stochastic process $v(t)$
is not ergodic. This also implies that
$\langle x^2(t)\rangle=v_T^2 t^2$, i.e. diffusion is asymptotically
ballistic, universally for any $0<\alpha<1$. 
Clearly, this
is an asymptotic regime of fully aged $K_v(t+t_{\rm ag},t_{\rm ag})$, 
$t_{\rm ag}\to\infty $.
It was not studied in Refs.  \cite{Barkai00,FriedrichPRL06,FriedrichPRE06}, 
where $t_{\rm ag}=0$
in fact.
Interestingly, in this latter case we obtain also 
$K_v(t,0)=v_T^2 E_\alpha(-r_\alpha t^\alpha)$, i.e. the same result as for
a different model in \cite{Barkai00}. From it, one cannot, however, conclude
anything on the behavior of $\langle x^2(t)\rangle$, 
even for $t_{\rm ag}=0$.
The correct asymptotical result in the case $t_{\rm ag}=0$ is
$\langle x^2(t)\rangle\sim (1-\alpha)v_T^2 t^2$, as 
it will be shown  below from a different perspective
in agreement with \cite{FriedrichPRL06,FriedrichPRE06} for the retarded version of 
the fractional
KFP equation. 
We stress, however, that a finite value of 
$\langle \tau \rangle$ is a very essential feature of our approach justifying 
$t_{\rm ag}=\infty$ as a very common experimental condition. Actually, it is hard
to imagine how an experimental system of scattering particles can be prepared
exactly at $t_{\rm ag}=0$.

\subsection{The limit $\langle \tau \rangle\to 0$, $r_\alpha\to \infty$, 
$\langle \tau \rangle r_\alpha =const$}

Another important limit is $\langle \tau \rangle=1/r\to 0$, 
$r_\alpha\to \infty$, i.e. of very fast scattering events, so that 
$\tau_r=(r_{\alpha}/r)^{1/(1-\alpha)}=const$.
 In this case,
$K_v^{\rm (st)}(t)=v_T^2 E_{1-\alpha}[-(t/\tau_r)^{1-\alpha}]$,
which  reminds 
the Barkai and Silbey result \cite{Barkai00} for $r\to 0$ and nonstationary 
$K_v(t,0)$. However, our result
contains $\alpha$ instead of $1-\alpha$, and also its meaning is very different,
in spite of a perplexedly confusing similarity.
Furthermore, 
a similar result, but with $2-\alpha$ instead of $1-\alpha$ and $1<\alpha<2$ was obtained
for the stationary velocity ACF within a very different model of super-diffusion
based on the fractional Langevin equation with super-Ohmic coupling to 
a thermal bath of harmonic oscillators 
\cite{MainardiPeroni,Lutz,SiegleEPL,GoychukACP12}. 
The physics of both models is, however, entirely
different. In our case,
the corresponding position variance grows as
\begin{eqnarray}\label{result1}
\langle x^2(t)\rangle= 2(v_Tt)^2E_{1-\alpha,3}[-(t/\tau_r)^{1-\alpha}],
\end{eqnarray}
where
$E_{\alpha,b}(z)=\sum_{n=0}^{\infty}z^n/\Gamma(\alpha n+b)$ is a generalized Mittag-Leffler
function. For $t\ll \tau_r$, $\langle x^2(t)\rangle\approx (v_Tt)^2$, and diffusion
is initially ballistic. This is because $K_v^{\rm (st)}(t)\approx v_T^2$ on this time scale.
For $t\gg \tau_r$, $K_v^{\rm (st)}(t)\propto 1/t^{1-\alpha}$, and 
$\langle x^2(t)\rangle\propto t^{1+\alpha}$, i.e. an asymptotic sub-ballistic 
superdiffusion
regime emerges. Interestingly, it is mostly close to the ballistic diffusion for
$\alpha=1-\epsilon$, $\epsilon\to +0$, and not for $\alpha\to 0$, as in Ref. \cite{Barkai00}.
This is a rather paradoxical and unexpected feature since in this case the anomalous
scattering channel is mostly close to the normal one within the model studied.

\subsection{General case of $\langle \tau\rangle>\tau_r$}

\begin{figure}
\resizebox{0.9\columnwidth}{!}{\includegraphics{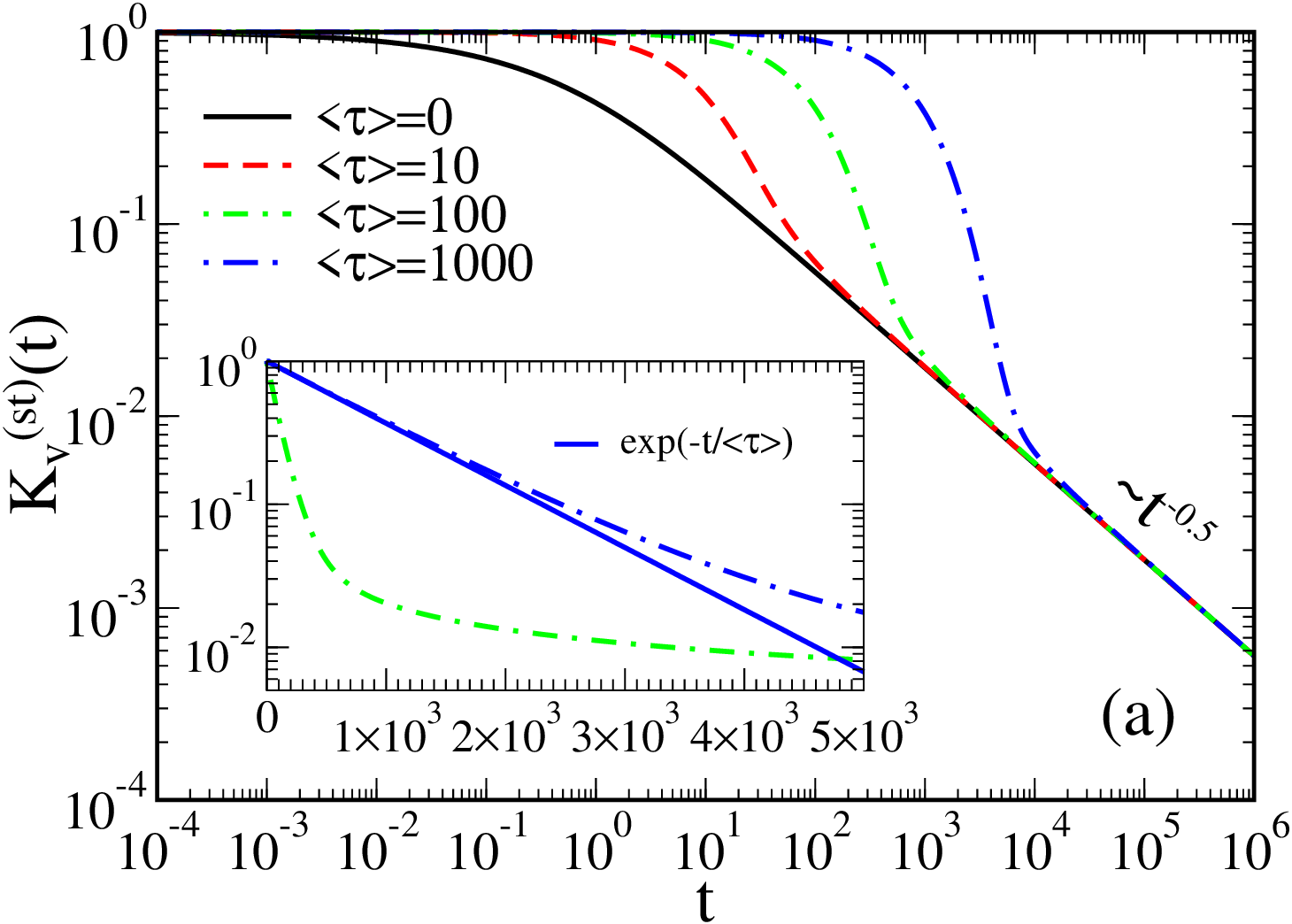}}
\resizebox{0.9\columnwidth}{!}{\includegraphics{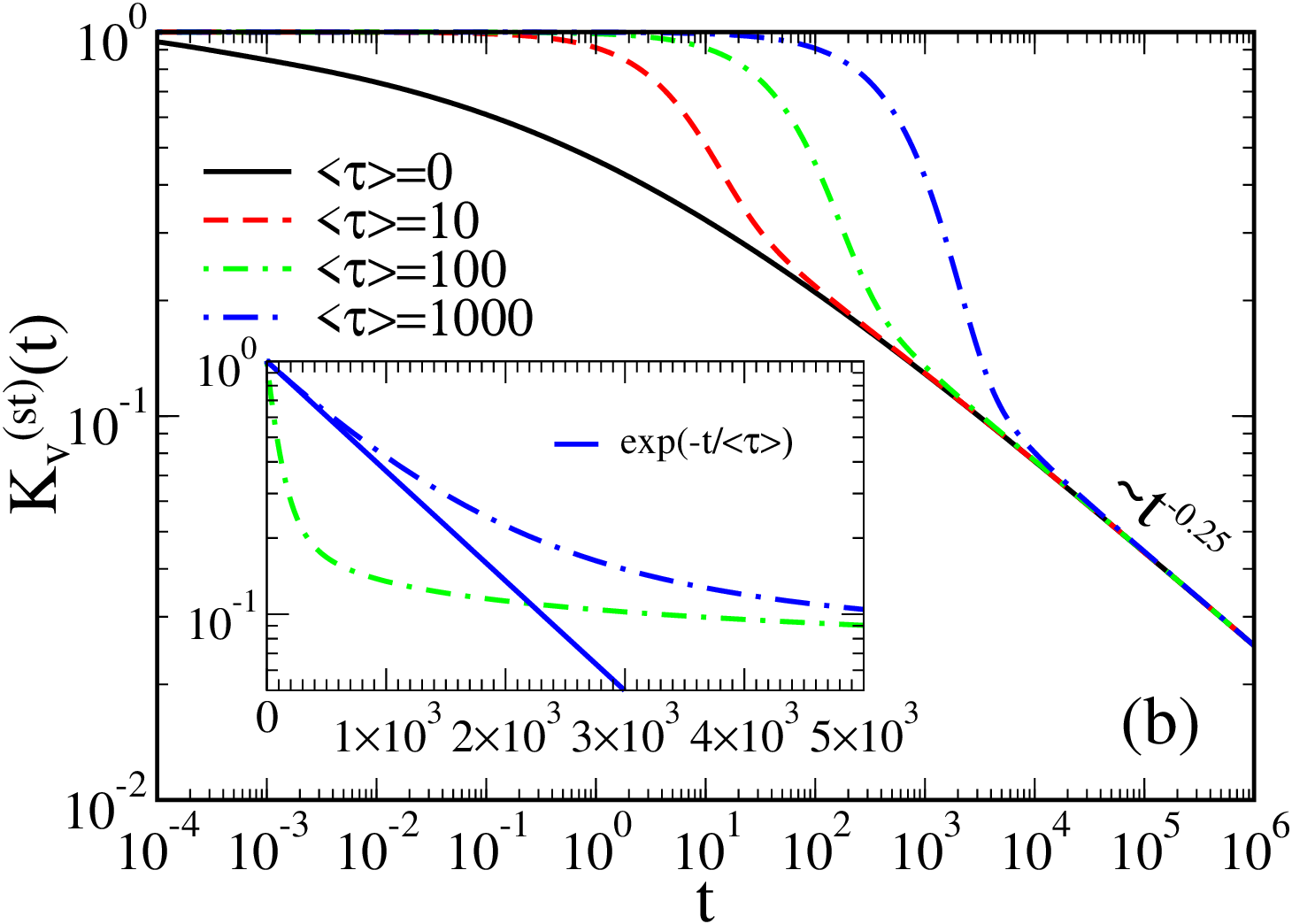}}
\caption{(Color online) Normalized stationary velocity autocorrelation function as function of time
(in units of $\tau_r$) for (a) $\alpha=0.5$, and (b) $\alpha=0.75$, as well as different
values of $\langle \tau\rangle$ shown in the plot. The results in (a) are exact analytical
results, whereas the results in (b) are obtained by a numerically precise inversion of the
corresponding Laplace transform using the Stehfest-Gaver algorithm.
The inset shows a part of the same plot on semi-logarithmic scale to
reveal an initially exponential decay indicated by a solid
line in the inset for $\langle \tau\rangle=1000$.}
\label{Fig1}       
\end{figure}

\begin{figure}
\resizebox{0.9\columnwidth}{!}{\includegraphics{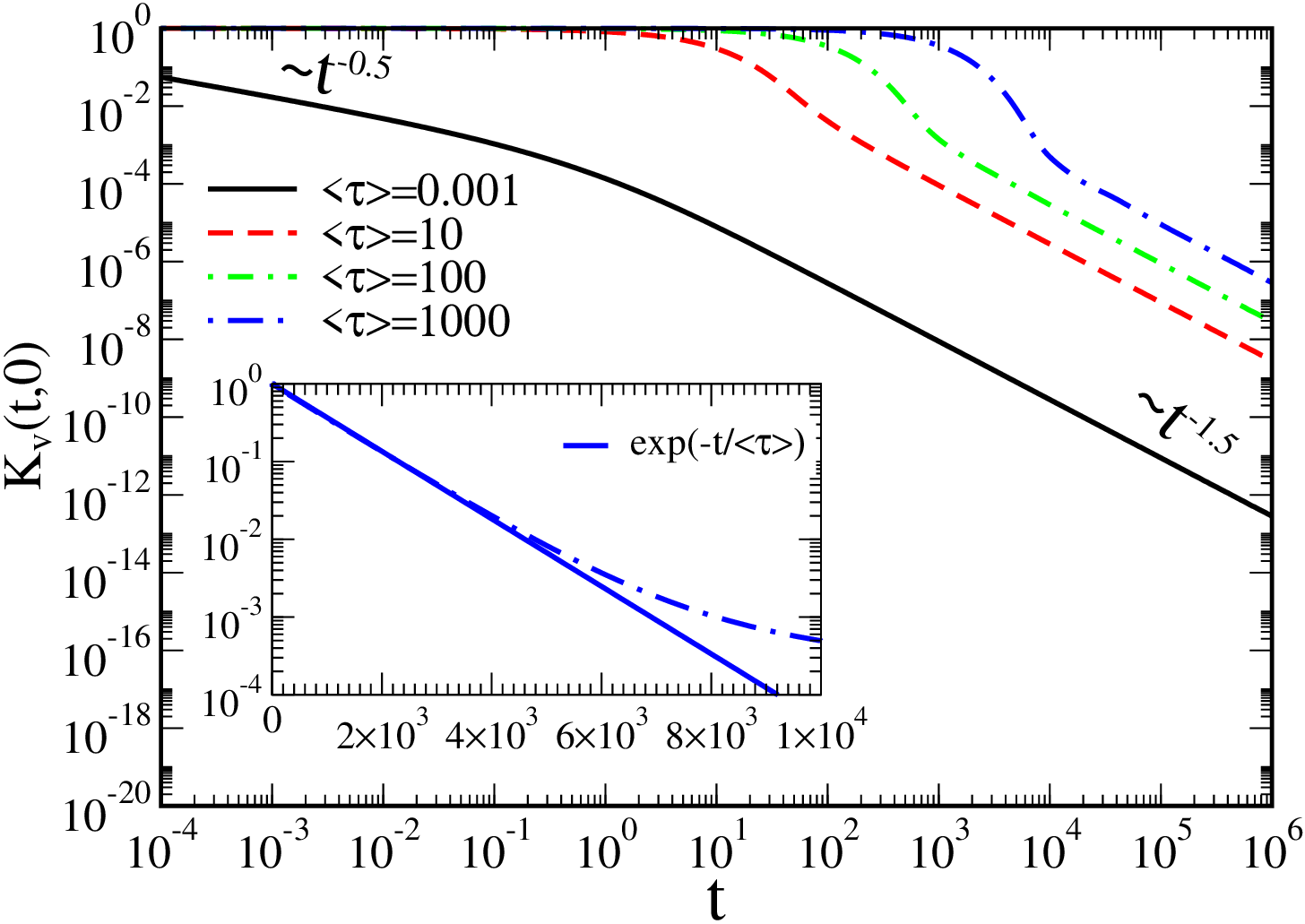}}
\caption{(Color online) Normalized velocity autocorrelation function $K_v(t,0)$ as function of time
(in units of $\tau_r$) for $\alpha=0.5$, and different
values of $\langle \tau\rangle$ shown in the plot. 
The inset shows a part of the same plot on semi-logarithmic scale to
reveal an initially exponential decay indicated by a solid
line in the inset for $\langle \tau\rangle=1000$. }
\label{Fig2}       
\end{figure}

The result of previous subsection holds approximately very good for 
$\langle \tau\rangle\ll \tau_r$. 
For $\langle \tau\rangle\gg\tau_r$, $K_v^{\rm (st)}(t)$
decays first exponentially, $K_v^{\rm (st)}(t)\approx v_T^2\exp(-t/\langle \tau\rangle)$,
and at $t_c\sim c\langle\tau\rangle$, where $c$ is a numerical coefficient about
$c\sim 10$, a transition occurs to an
algebraic tail behavior, $K_v^{\rm (st)}(t)\propto 1/t^{1-\alpha}$, 
see in Fig. \ref{Fig1}. Therein, $K_v^{\rm (st)}(t)$ is plotted for different values
of $\langle \tau\rangle$ in the cases $\alpha=1/2$, where an exact analytical expression
can be readily found by the inversion of the Laplace-transform, and $\alpha=0.75$,
where we invert the Laplace-transform numerically using the Stehfest-Gaver algorithm \cite{Stehfest}. 
The 
analytical expression for $\alpha=0.5$ reads
\begin{eqnarray}\label{Kv_st_exact}
&& K_v^{\rm (st)}(t)=\frac{v_T^2}{\sqrt{\tau_1}-\sqrt{\tau_2}}\\
&& \times 
\left [\sqrt{\tau_1} E_{1/2} \left (-\sqrt{\frac{t}{\tau_1}} \right ) 
-\sqrt{\tau_2} E_{1/2} \left (-\sqrt{\frac{t}{\tau_2}} \right ) 
  \right ]\;,\nonumber
\end{eqnarray}
where $E_{1/2}(-\sqrt{z})=\exp(z){\rm erfc}(\sqrt{z})$, is the Mittag-Leffler
function with index $\alpha=1/2$ and argument $-\sqrt{z}$, and 
$\tau_{1,2}=\tau_r\left (1\pm\sqrt{1-4\langle \tau\rangle/\tau_r}\right )^2/4$.
The stationary ACF has a power law tail, $K_v^{\rm (st)}(t)\propto t^{-1+\alpha}$, see in 
Fig. \ref{Fig1}.
A similar algebraic tail features also the velocity ACF
in simple fluids. It emerges due to the hydrodynamic memory effects
yielding $K_v^{\rm (st)}(t)\propto t^{-1.5}$ asymptotically 
\cite{Zwanzig,MainardiPeroni,GoychukACP12}. 
However, very
different from the hydrodynamic memory
 case, in our model this algebraic tail is not integrable
and it yields asymptotically a superdiffusion, 
$\langle x^2(t)\rangle\propto t^{1+\alpha}$. In this respect,
 it is also very important to mention that the
zero-age $K_v(t,0)$ for $\alpha=1/2$ reads
\begin{eqnarray}\label{Kv0_exact}
&& K_v(t,0)=\frac{v_T^2}{\sqrt{\tau_1}-\sqrt{\tau_2}}\\
&& \times 
\left [\sqrt{\tau_1} E_{1/2} \left (-\sqrt{\frac{t}{\tau_2}} \right ) 
-\sqrt{\tau_2} E_{1/2} \left (-\sqrt{\frac{t}{\tau_1}} \right ) 
  \right ]\;.\nonumber
\end{eqnarray}
Notice a very subtle difference between (\ref{Kv0_exact}) and (\ref{Kv_st_exact}), which is
not easy to spot!
A structurally very same equation was obtained for the \textit{stationary} velocity ACF
within the fractional Langevin equation by Mainardi and Peroni \cite{MainardiPeroni}, wherein $r$
is related to the standard Stokes friction and $r_{1/2}$ to the hydrodynamic
memory effects. Asymptotically, $K_v(t,0)\propto t^{-3/2}$.
For a large $\langle \tau \rangle$, it starts also from an exponential part
$K_v(t,0)\approx \exp(-t/\langle \tau \rangle)$, like $K_v^{\rm (st)}(t)$, cf. 
in Fig. \ref{Fig2}, which ends in the $t^{-3/2}$ tail.
 For a small $\langle \tau \rangle$,
it displays also another intermediate power law $
K_v(t,0)\propto t^{-1/2}$, like  $K_v^{\rm (st)}(t)$ asymptotically.

Furthermore, the exact expression for
the position variance in the case $\alpha=1/2$
reads
\begin{eqnarray}\label{diffusion_exact}
 &&\langle x^2(t)\rangle =2(v_T t)^2 \frac{1}{\sqrt{\tau_1}-\sqrt{\tau_2}}\\
&& \times 
\left [\sqrt{\tau_1} E_{1/2,3} \left (-\sqrt{\frac{t}{\tau_1}} \right ) 
-\sqrt{\tau_2} E_{1/2,3} \left (-\sqrt{\frac{t}{\tau_2}} \right ) 
  \right ]\;,\nonumber 
\end{eqnarray}
where 
\begin{eqnarray}
E_{1/2,3} (-\sqrt{z})=z^{-2} \Big [ -1+ \frac{2}{\sqrt{\pi}} z^{1/2}-z
+\frac{4}{3\sqrt{\pi}}z^{3/2}\nonumber \\
+E_{1/2} (-\sqrt{z} )\Big ]\;.
\end{eqnarray}

\begin{figure}
\resizebox{0.9\columnwidth}{!}{\includegraphics{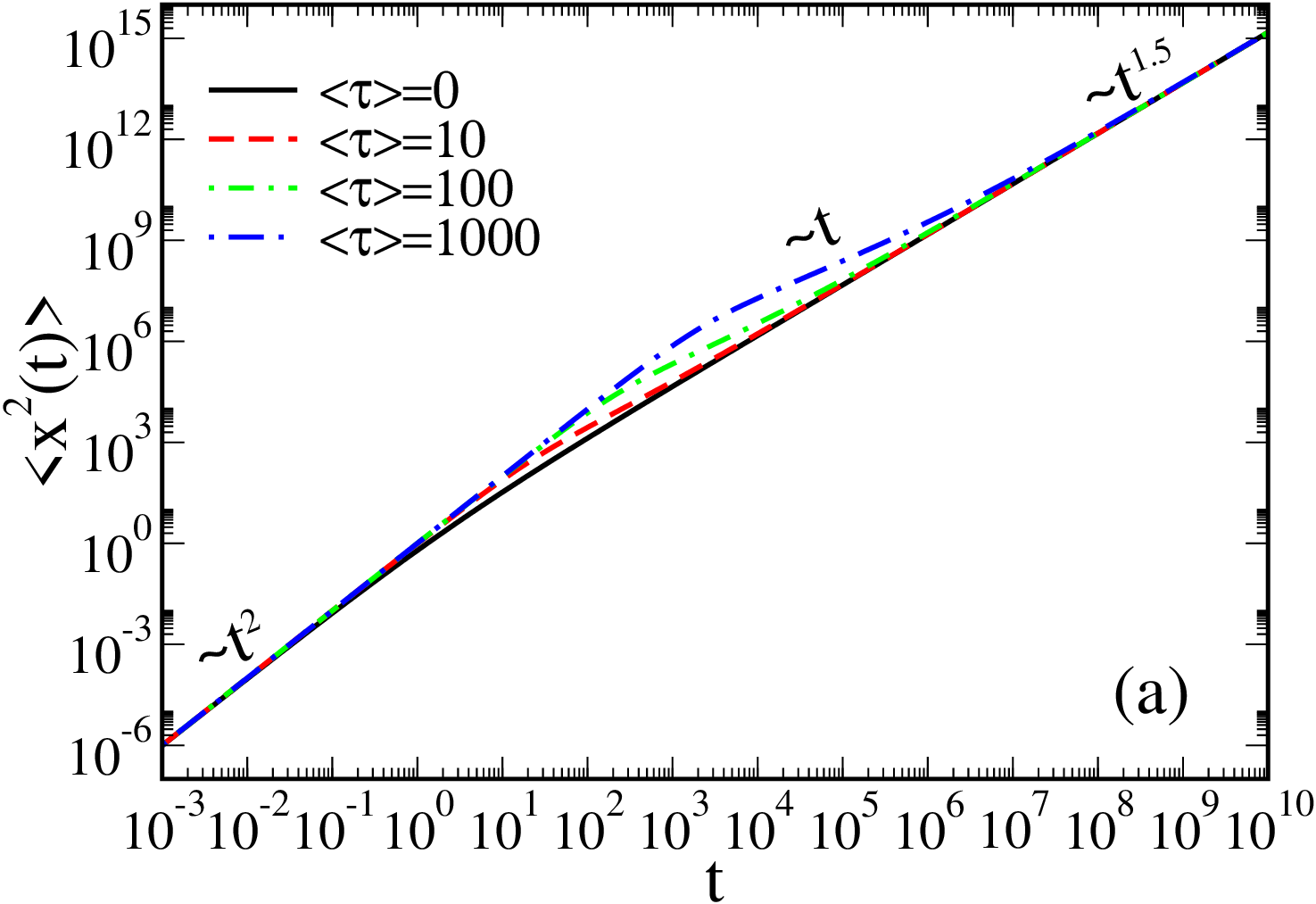}}
\resizebox{0.9\columnwidth}{!}{\includegraphics{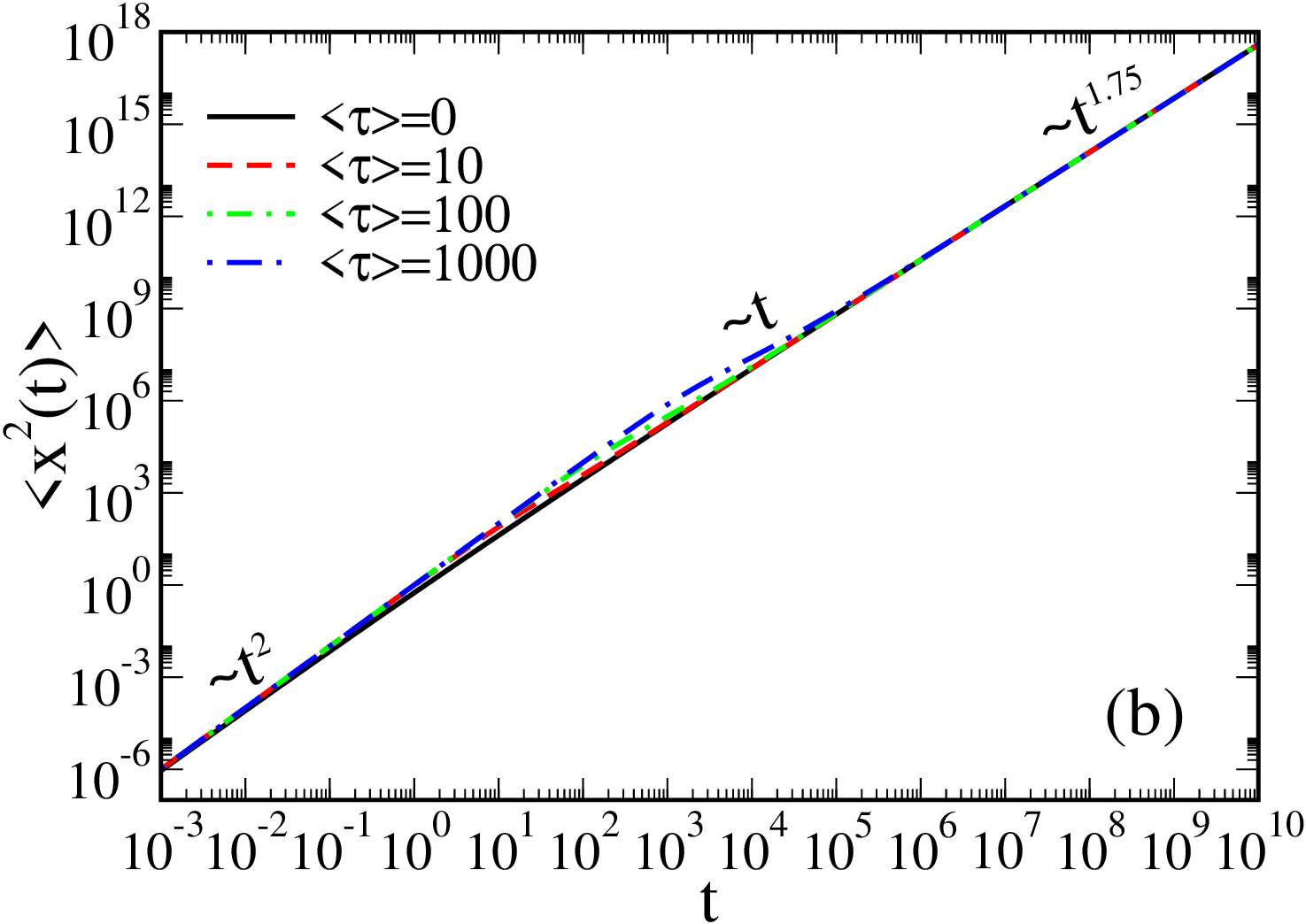}}
\caption{(Color online) Variance of particles position (in units of $(v_T\tau_r)^2$) 
as function of time
(in units of $\tau_r$) for (a) $\alpha=0.5$, and (b) $\alpha=0.75$, as well as different
values of $\langle \tau\rangle$ shown in the plot. The results in (a) are exact analytical
results, while the results in (b) are obtained by a numerically precise inversion of the
corresponding Laplace transform.}
\label{Fig3}       
\end{figure}

This expression shares the following general features for other
values of $\alpha$. For $\langle \tau\rangle\gg\tau_r$,
diffusion is initially ballistic for $0<t\ll \langle \tau\rangle$. Then, it becomes
transiently normal for $\langle \tau\rangle <t < c \langle \tau\rangle$.
Finally, after slowing down it again accelerates and becomes sub-ballistic
superdiffusion with 
\begin{eqnarray}\label{asymp_res1}
\langle x^2(t)\rangle\sim 2(v_T\tau_r)^2 (t/\tau_r)^{1+\alpha}/\Gamma(2+\alpha)
\end{eqnarray}
for $t\gg c \langle \tau\rangle$, see in Fig. \ref{Fig3}. Notice that
this is the same asymptotics independently of $\langle \tau\rangle$ as
one produced by the result in Eq. (\ref{result1}).
One can clearly see that with growing $\langle \tau\rangle$, the initial regime
of ballistic diffusion extends gradually 
to infinity while $\langle \tau\rangle\to\infty$, independently
of $\alpha$. 

\begin{figure}
\resizebox{0.9\columnwidth}{!}{\includegraphics{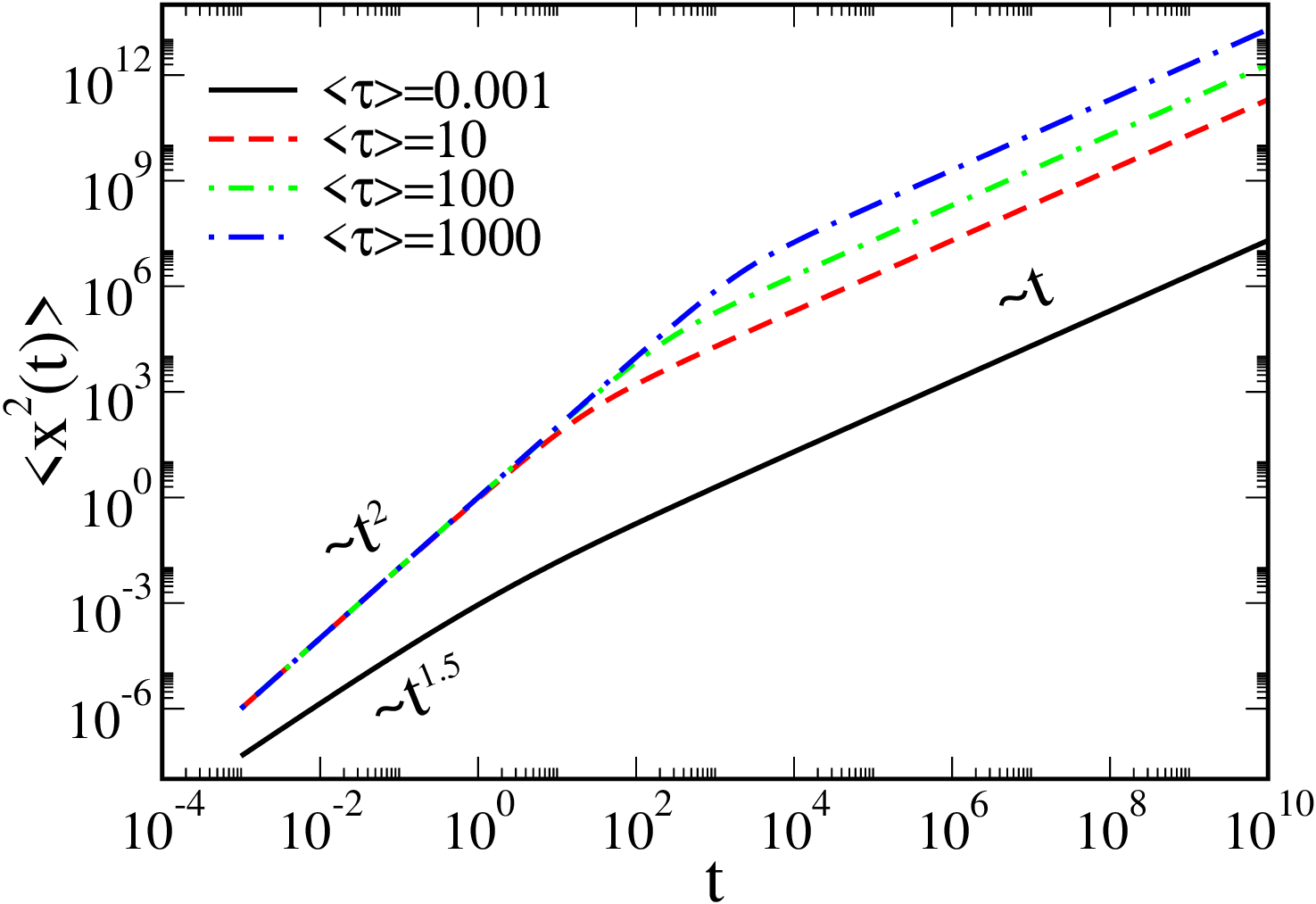}}
\caption{(Color online) Variance of particles position (in units of $(v_T\tau_r)^2$) as function of time
(in units of $\tau_r$) for $\alpha=0.5$ and different values of $\langle \tau\rangle$ 
if to substitute \textit{ad hoc} 
$K_v^{\rm (st)}(t)$ with $K_v(t,0)$ in Eq. (\ref{diff1}).}
\label{Fig4}       
\end{figure}

Furthermore, if to use mistakingly $K_v(t,0)$ instead of  $K_v^{\rm (st)}(t)$ 
in Eq. (\ref{diff1}), we obtain
\begin{eqnarray}\label{diffusion_wrong}
 &&\langle x^2(t)\rangle =2(v_T t)^2 \frac{1}{\sqrt{\tau_1}-\sqrt{\tau_2}}\\
&& \times 
\left [\sqrt{\tau_1} E_{1/2,3} \left (-\sqrt{\frac{t}{\tau_2}} \right ) 
-\sqrt{\tau_2} E_{1/2,3} \left (-\sqrt{\frac{t}{\tau_1}} \right ) 
  \right ]\;.\nonumber 
\end{eqnarray}
The formal difference with (\ref{diffusion_exact}) is not easy to detect.
However, the diffusive behavior is very different, see in Fig. \ref{Fig4}
and compare with Fig. \ref{Fig3}, (a). First of all, asymptotically
this is a normal diffusion, although the initial regime of ballistic diffusion, which
is also universal, gradually extends to infinity with growing $\langle \tau\rangle$.
Intermittently, it can be sub-ballistic superdiffusion with 
$\langle x^2(t)\rangle\propto t^{3/2}$, as for $\langle \tau\rangle=0.001$ 
in Fig. \ref{Fig4} (it looks like initial regime therein because the 
truly initial ballistic
regime is simply not depicted for this value of parameter).

\section{Fractional BGK equations in the phase space}

We proceed further with a generalization of the above description from the velocity
subspace to the whole phase space because we wish to have a kinetic description valid
in arbitrary force fields $F\neq 0$.
It turns out to be a very nontrivial task.
First, following Friedrich {\it et al.} \cite{FriedrichPRL06,FriedrichPRE06} one 
must take the retardation effects into
account. For the case $F=0$, this can be done exactly. Namely,
from Eq. (\ref{GME}) we obtain the corresponding scattering integral
\begin{eqnarray}\label{memory1}
&&{\rm St}[f(x,v,t)]= -r(v)\int_0^{t} dt'K(t-t')f(X(t,t'),v,t')\nonumber \\
&& +\int_0^{t}  dt'K(t-t')\int dv'w(v\leftarrow v')f(X'(t,t'),v',t')\nonumber
\end{eqnarray} 
in the phase space,
where $X(t,t')$  and $X'(t,t')$ in Eq. (\ref{memory1}) are the retarded spatial
variables reading, $X(t,t')=x-v(t-t')$ and $X'(t,t')=x-v'(t-t')$.  Because $F=0$, the velocity
remains constant between two scattering events. The physical meaning of these 
variables and their origin is clear. 
Because $x(t)=\int_0^t v(t')dt'$, both 
$X(t,t')$ and $X'(t,t')$ are just the values of the coordinate $x$ at the time 
$t'$, i.e. $x(t')$, for two different constant values of the velocity variable. 
This retardation, which results
from a rigorous treatment of the scattering process 
in the phase space \cite{FriedrichPRL06,FriedrichPRE06}, 
makes everything rather intricate,
especially in the presence of external force fields, 
when $F\neq 0$. Then, most obviously not only $X(t,t')=x-\int_{t'}^t v(\tau)d\tau$ generally,
but also a retardation 
in the velocity variable should be taken into account. To neglect this latter one
is only possible if the acceleration of the particles between two scattering
events is negligible.
We wish to explore a possibility to avoid such
complexities and to disregard the retardation effects whenever it can safely be done. 
Intuitively, this can only be physically justified
if $\langle \tau\rangle$ is finite and a relative change of $f(x,v,t)$ on the spatial scale
$x_T=v_T\langle \tau\rangle$ is completely negligible.
At the first look, it seems that
using $\langle \tau\rangle$ sufficiently small, this approximation can always be justified,
which is definitely not the case of $\langle \tau\rangle=\infty$, like in the case of fractional 
Kramers-Fokker-Planck equation by Barkai and Silbey. 
The latter one corresponds to the case, 
where the only one anomalous
scattering channel is present and the scattering integral is taken in the KFP form
with a velocity-independent friction coefficient $\eta$, and the retardation
effects are completely disregarded.
For the considered model of scattering mechanism, a general fractional linear Boltzmann
equation (FLBE) with retardation can readily be written
\begin{eqnarray}\label{FLBE}
&&\frac{\partial f(x,v,t)}{\partial t}+
v\frac{\partial f(x,v,t)}{\partial x}+\frac{F(x,v,t)}{M}
\frac{\partial f(x,v,t)}{\partial v}\nonumber \\
=&&-\left (1+(r_{\alpha}/r)\sideset{_0}{_t}{\mathop{\hat
{\cal D}}^{1-\alpha_j}}\right )\Big [r(v)f(x,v,t')\\
&&-\int dv'w(v\leftarrow v')f(x,v',t')\Big ], \nonumber
\end{eqnarray}
where 
\begin{eqnarray}\label{substantialRL} 
&&\sideset{_{0}}{_t}{\mathop{\hat {\cal D}}^{\gamma}}f(x,v,t)=\frac{1}{\Gamma(1-\gamma)}\nonumber \\ 
&&\times \left(
\frac{\partial}{\partial t}+v\frac{\partial}{\partial x}\right )\int_{0}^t dt' \frac{f(x-v(t-t'),v,t')}{(t-t')^\gamma}\;.
\end{eqnarray}
is the operator of substantial fractional derivative, which takes the retardation of
the $x$ variable into account \cite{FriedrichPRL06,FriedrichPRE06}.
Here, one implicitly assumes that between two scattering events
the particles velocity is not changed in spite of the external force applied.
This, of course, can only be done if either $\langle \tau\rangle$ is sufficiently small,
or when $F=0$ exactly.
Furthermore, a fractional generalization of the 
BGK kinetic equation or fractional BGK kinetic equation (FBGKE) reads
\begin{eqnarray}\label{FBGE}
&&\frac{\partial f(x,v,t)}{\partial t}+
v\frac{\partial f(x,v,t)}{\partial x}+\frac{F(x,v,t)}{M}
\frac{\partial f(x,v,t)}{\partial v}\\
=&&-\left (r+r_{\alpha}
\sideset{_0}{_t}{\mathop{\hat 
{\cal D}}^{1-\alpha}}\right )\left [f(x,v,t)-f_M(v)\int_{}^{} dv f(x,v,t) \right ]\;.
\nonumber
\end{eqnarray}

These kinetic equations correspond, however, to the case when the scattering
events started at $t_0=0$. What will happen if the scattering events started in the infinite past
at the age $-t_{\rm ag}$, $t_{\rm ag}\to\infty$? 
In this case, we can adopt GME (\ref{GME2}) for writing the collision term upon
taking the retardation effects in the $x$ variable into account in the following
manner:
\begin{eqnarray}\label{memory2}
&&{\rm St}[f(x,v,t)]= -r(v)\int_0^{t} dt'K_{\alpha}(t-t')\\
&&\times [f(X(t,t'),v,t')-f(X(t,0),v,0) ]\nonumber \\
&& +\int_0^{t}  dt'K_{\alpha}(t-t')\int dv'w(v\leftarrow v')\nonumber \\
&&\times [f(X'(t,t'),v',t')-f(X'(t,0),v',0)] \nonumber \\
&&-r(v)f(x,v,t)+\int dv' w(v\leftarrow v') f(x,v',t)\;.\nonumber
\end{eqnarray} 
Here, a singular memory kernel (which is not a function but a distribution) 
has the Laplace-transform 
$\tilde K_\alpha(s)=r_{\alpha}s^{1-\alpha}$.

The corresponding fully-aged version of our fractional BGK equation with retardation reads
\begin{eqnarray}\label{FBGE2}
&&\frac{\partial f(x,v,t)}{\partial t}+
v\frac{\partial f(x,v,t)}{\partial x}+\frac{F(x,v,t)}{M}
\frac{\partial f(x,v,t)}{\partial v} \\
=&&-\Big (r+r_{\alpha}
\sideset{_0}{_t}{\mathop{\hat 
{\cal D}}^{1-\alpha}} \Big )\Big[ f(x,v,t) - f_M(v)\int dv f(x,v,t)\Big ]\nonumber \\
&&+\int_0^tK_{\alpha}(t-t')\Big [f(x+vt,v,0)\nonumber \\
&&-f_M(v)
\int_{}^{}f(x+v't,v',0) dv'   \Big ]dt'\;.\nonumber 
\end{eqnarray}
Can we disregard the retardation overall and to replace the substantial fractional derivative
by the standard one, for example, in the limit $\langle \tau\rangle\to 0$? 
This is a fundamental question which will be answered below. 
Fractional kinetic equations (\ref{FLBE}), (\ref{FBGE}), and (\ref{FBGE2}) present
the central theoretical proposals of this paper.

\section{Fractional superdiffusion within FBGKE}

The next important task is to establish if we do can neglect the retardation effects
in the fractional kinetic equations in the phase space, and when 
it is possible in principle.

\subsection{Standard form of FBGKE with retardation and without}

We start from the standard form of 
the force-free ($F=0$) FBGKE (\ref{FBGE}), which takes the retardation effects into account. 
In terms of the double Fourier transform of the distribution function, \\
$\hat G(k,\eta,t)=\int_{-\infty}^{\infty}dx\int_{-\infty}^{\infty}dv e^{i(kx+\eta v)}f(x,v,t)$,
which is the moment-generating function,
  this equation can be written as
\begin{eqnarray}\label{FBGEgen1}
&&\frac{\partial \hat G(k,\eta,t)}{\partial t}-k \hat G'_{\eta}(k,\eta,t)=
-\int_0^t K(t-t') \\
&& \times \Big [ \hat  G(k,\eta+k(t-t'),t')
-e^{-(\eta v_T)^2/2}\hat  G(k,k(t-t'),t') \Big ],\nonumber
\end{eqnarray}
where $\hat G'_{\eta}(k,\eta,t)$ denotes a partial 
derivative with respect to the variable $\eta$,
with retardation effects taken into account, and 
\begin{eqnarray}\label{FBGEgen2}
&&\frac{\partial \hat G(k,\eta,t)}{\partial t}-k \hat G'_{\eta}(k,\eta,t)=
-\int_0^t K(t-t') \\
&& \times \Big [ \hat  G(k,\eta,t')
-e^{-(\eta v_T)^2/2}\hat  G(k,0,t') \Big ],\nonumber
\end{eqnarray}
without.
These equations are difficult to solve. 
However, by setting $k=0$ therein we can readily deduce that the corresponding
FBGE in the velocity space reads (\ref{v-space}) independently of whether we took the retardation
effects  into account or not. This is the same feature as with 
the fractional KFP
equation \cite{FriedrichPRE06}. Furthermore, the discussed equations can be used to find the equations of motion for 
the moments of coordinate and velocity by taking a corresponding number of derivatives of 
$ \hat G(k,\eta,t)$ with respect to $k$ and $\eta$ at $k=0,\eta=0$. In this way, we obtain
from Eq. (\ref{FBGEgen1}):
\begin{eqnarray}\label{moment1}
&&\frac{d}{dt}\langle x^2(t)\rangle =2\langle x(t)v(t)\rangle,\\
&&\frac{d}{dt}\langle x(t)v(t)\rangle=\langle v^2(t)\rangle-\int_0^t K(t-t')\label{moment2} \\
&& \times\Big [ \langle x(t')v(t')\rangle +\langle v^2(t')\rangle(t-t')\Big ]dt',\nonumber \\
&&\frac{d}{dt}\langle v^2(t)\rangle=-\int_0^t K(t-t') \label{moment3}
\big [ \langle v^2(t')\rangle- v_T^2\big ]dt'\,.
\end{eqnarray} 
If to neglect the retardation effects, Eq. (\ref{moment2}) is replaced by
\begin{eqnarray}\label{moment2mod}
\frac{d}{dt}\langle x(t)v(t)\rangle=\langle v^2(t)\rangle-\int_0^t K(t-t') 
\langle x(t')v(t')dt'.
\end{eqnarray}
This is the only difference. Solving Eq. (\ref{moment3}), we obtain
\begin{eqnarray}
\widetilde{\langle v^2(s)\rangle}=\frac{\langle v^2(0)\rangle s+v_T^2\tilde K(s) }{s[s+\tilde K(s)]},
\end{eqnarray}
or $\langle v^2(t)\rangle =v_T^2+\Phi(t) [\langle v^2(0)\rangle-v_T^2]$ in the time domain, which
is consistent with the above-given solution for $p(v,t)$ in this case. The latter one has precisely 
the same relaxation structure. Furthermore, Eq. (\ref{moment2}) can be solved
by using the convolution theorem and noticing that the Laplace-transform of $K(t)t$ reads $-\tilde K'(s)$,
where $\tilde K'(s)$ is the derivative over $s$.
Finally, with the initial conditions $\langle x^2(0)\rangle=0$ and  $\langle x(0)v(0)\rangle=0$
we obtain for the Laplace-transformed position variance
\begin{eqnarray}
\widetilde{\langle x^2(s)\rangle}=
\frac{2[1+\tilde K'(s)][\langle v^2(0)\rangle s +v_T^2\tilde K(s)] }{s^2[s+\tilde K(s)]^2}\;.
\end{eqnarray}
Similar expression for a particular case $\langle v^2(0)\rangle =0$ has 
been obtained by Friedrich \textit{et al.}
\cite{FriedrichPRE06} for a different model (fractional Kramers-Fokker-Planck equation
with retardation effects) in different notations. To compare with the above
solutions obtained in the velocity domain, it is useful to take the  equilibrium distribution
of velocities initially. This yields
\begin{eqnarray}\label{41}
\widetilde{\langle x^2(s)\rangle}=
\frac{2v_T^2[1+\tilde K'(s)] }{s^2[s+\tilde K(s)]}\;.
\end{eqnarray}
Furthermore, if to neglect the retardation effects the latter equation is modified as
\begin{eqnarray}\label{wrong}
\widetilde{\langle x^2(s)\rangle}=
\frac{2v_T^2 }{s^2[s+\tilde K(s)]}\;.
\end{eqnarray}
It gives precisely the same result in the time domain as if incorrectly 
use $K_v(t,0)$ in Eq. (\ref{diff1}) instead of $K_v^{\rm (st)}(t)$. This is
actually very misleading! This masking unfortunate feature reflects 
profound problems emerging immediately
if to neglect the retardation effects.
Let us discuss the related subtleties.

\subsubsection{Fractional FBGKE with infinite $\langle \tau\rangle$}

In this case $r=0$, and Eq. (\ref{41}) yields
\begin{eqnarray}
\widetilde{\langle x^2(s)\rangle}=
\frac{2v_T^2[1+(1-\alpha)r_\alpha s^{-\alpha}] }{s^2[s+r_\alpha s^{1-\alpha}]}\;,
\end{eqnarray}
including retardation effects. The corresponding \\
$\langle x^2(t)\rangle=
\langle x^2_1(t)\rangle+\langle x^2_2(t)\rangle$ consists
of two parts. The first corresponds to the neglect
of the retardation effects, and this is precisely the result by Barkai
and Silbey, 
$\langle x_1^2(t)\rangle= 2(v_Tt)^2E_{\alpha,3}[-r_\alpha t^{\alpha}]$, 
obtained for a different model. Asymptotically, 
$\langle x_1^2(t)\rangle\propto t^{2-\alpha}$. Notice, however,
that in the sum  the second term, 
$\langle x_2^2(t)\rangle\sim v_T^2 (1-\alpha) t^2$, dominates asymptotically, 
which is the result by
Friedrich \textit{et al.} for a different KFP model \cite{FriedrichPRE06}. 
Interestingly, the same asymptotics and a similar subleading term 
were also obtained by Barkai and Fleurov within a kinetically 
related model \cite{Barkai96}. The same ballistic asymptotics
with prefactor $1-\alpha$ was obtained also by Zumofen and Klafter
within a different model \cite{ZumofenPhysicaD}.
This result is the correct
result for the case of zero age, $t_{\rm ag}=0$. Furthermore,
notice the difference of prefactors $1-\alpha$ for $t_{\rm ag}=0$, and
just one for $t_{\rm ag}=\infty$, where $K_v^{\rm (st)}(t)=v_T^2=const$.

\subsubsection{Fractional FBGKE with finite  $\langle \tau\rangle$} 
 
In this case,  
\begin{eqnarray}
\widetilde{\langle x^2(s)\rangle}=
\frac{2v_T^2[1+(1-\alpha)r_\alpha s^{-\alpha}] }{s^2[s+r+r_\alpha s^{1-\alpha}]}\;.
\end{eqnarray} 
Notice the difference with the result obtained by using $K_v^{\rm (st)}(t)$
in Eq. (\ref{diff1}),
which can be obtained from the above expression 
by replacing $1-\alpha$ with unity in a prefactor in the numerator.
This leads to the result that asymptotically diffusion is slower by the factor of $1-\alpha$
than one given in Eq. (\ref{asymp_res1}) and 
depicted in Fig. \ref{Fig3}. Nevertheless, the asymptotic behavior, 
$\langle x^2(t)\rangle\propto t^{1+\alpha}$ is qualitatively correctly reproduced, as well
as the regime of initially ballistic diffusion. However, if we neglect the
retardation effects we obtain again the result 
which would correspond to the use of $K_v(t,0)$ in Eq. (\ref{diff1}) 
instead of $K_v^{\rm (st)}(t)$. It displays 
a completely wrong asymptotical behavior, namely
a normal diffusion, as depicted in Fig. \ref{Fig4} for $\alpha=1/2$.
Notice that this profound mistake of approximation
 persists even in the limit $\langle \tau\rangle\to 0$!
Hence, the intuition misleads and one cannot neglect the retardation
effects in the fractional FBGKE dynamics even in this limit. This
is contrary to our initial expectations. A proper mathematical treatment
defeats intuition.

\subsection{FBGKE with retardation and time-shift invariant scattering integral}

Finally, we would like to clarify whether
our second FBGKE possessing the time-shift invariant scattering
term does yield the correct results for diffusion obtained earlier from $K_v^{\rm (st)}(t)$. 
This is a very important self-consistency test.
In terms of $\hat G(k,\eta,t)$, Eq. (\ref{FBGE2}) 
 can be written as
\begin{eqnarray}\label{FBGE2gen}
&&\frac{\partial \hat G(k,\eta,t)}{\partial t}-k \hat G'_{\eta}(k,\eta,t)=
-\int_0^t K_{\alpha}(t-t') \\
&&\times \big [ \hat  G(k,\eta+k(t-t'),t')-e^{-(\eta v_T)^2/2}\hat  G(k,k(t-t'),t')\nonumber \\
&&-\hat  G(k,\eta+kt,0)+e^{-(\eta v_T)^2/2}\hat  G(k,kt,0)
\big ]dt' \nonumber \\
&&-r \Big [\hat  G(k,\eta,t)-e^{-(\eta v_T)^2/2} \hat  G(k,0,t) \Big ]\;.\nonumber
\end{eqnarray}
Furthermore, if to neglect the retardation, it becomes
\begin{eqnarray}\label{FBGE3gen}
&&\frac{\partial \hat G(k,\eta,t)}{\partial t}-k \hat G'_{\eta}(k,\eta,t)=
-\int_0^t K_{\alpha}(t-t') \\
&&\times \big [ \hat  G(k,\eta,t')-e^{-(\eta v_T)^2/2}\hat  G(k,0,t')\nonumber \\
&&-\hat  G(k,\eta,0)+e^{-(\eta v_T)^2/2}\hat  G(k,0,0)
\big ]dt' \nonumber \\
&&-r \Big [\hat  G(k,\eta,t)-e^{-(\eta v_T)^2/2} \hat  G(k,0,t) \Big ].\nonumber
\end{eqnarray}
Using (\ref{FBGE2gen}), we obtain 
\begin{eqnarray}\label{moment2mod2}
&&\frac{d}{dt}\langle x(t)v(t)\rangle=\langle v^2(t)\rangle-\int_0^t K_{\alpha}(t-t') \\
&& \times\Big \{  \langle x(t')v(t')\rangle-\langle x(0)v(0)\rangle
 +\langle v^2(t')\rangle(t-t')-\langle v^2(0)\rangle t \Big \}dt'\nonumber \\
&& -r\langle x(t)v(t)\rangle \nonumber
\end{eqnarray}
instead of Eq. (\ref{moment2}) and
\begin{eqnarray}\label{moment3mod}
&&\frac{d}{dt}\langle v^2(t)\rangle=-\int_0^t K_{\alpha}(t-t')
\big [ \langle v^2(t')\rangle- \langle v^2(0)\rangle \big ]dt'\nonumber \\
&& -r \big [ \langle v^2(t)\rangle- v_T^2 \big ]
\end{eqnarray}
instead of Eq. (\ref{moment3}). Eq. (\ref{moment1}) remains, of course, always valid.
Furthermore, if to neglect retardation, Eq. (\ref{moment3mod}) remains
the same. Its solution reads \\
$\langle v^2(t)\rangle=
v_T^2+\Phi^{(0)}(t)[\langle v^2(0)\rangle -v_T^2]$, as expected from
the general velocity relaxation law within this model. However, Eq. (\ref{moment2mod2})
is get modified as
\begin{eqnarray}\label{moment2mod3}
&&\frac{d}{dt}\langle x(t)v(t)\rangle=\langle v^2(t)\rangle-\int_0^t K_{\alpha}(t-t') \\
&& \times \big [ \langle x(t')v(t')\rangle- \langle x(0)v(0)\rangle \big ]dt' 
-r\langle x(t)v(t)\rangle \;.\nonumber
\end{eqnarray}
This allows to immediately realize  that the neglect of retardation effects
yields asymptotically for $\langle v^2(0)\rangle=v_T^2$, $\langle x(0)\rangle v(0)\rangle=0$ the 
same incorrect result (\ref{wrong}). Hence, the
retardation effects are indispensable indeed, even in the limit $\langle \tau\rangle \to 0$,
within the considered fractional dynamics. With retardation effects
taken into account, we obtain for the initial conditions
$\langle x^2(0)\rangle=0, \langle x(0)v(0)\rangle=0$ the following remarkable result
\begin{eqnarray}\label{exact}
&&\widetilde{\langle x^2(s)\rangle}=
\frac{2v_T^2[1+\tilde K_{\alpha}(s)/s] }{s^2[s+r+\tilde K_{\alpha}(s)]} \\
&&+\frac{2(\langle v(0)^2\rangle- v_T^2 )}{s^2[s+r+\tilde K_{\alpha}(s)]} \nonumber \\
&& \times \Big [s \tilde \Phi^{(0)}(s)[1+\tilde K'_{\alpha}(s)]+
\tilde K_{\alpha}(s)/s-\tilde K'_{\alpha}(s) \Big ]\nonumber
\;.
\end{eqnarray}
It is valid for any memory kernel, which can be splitted
as $\tilde K(s)=r+\tilde K_{\alpha}(s)$, with $\tilde K_{\alpha}(0)=0$.
For the considered case of $\tilde K_{\alpha}(s)=r_{\alpha}s^{1-\alpha}$
we obtain
\begin{eqnarray}\label{exact2}
&&\widetilde{\langle x^2(s)\rangle}=
\frac{2v_T^2[1+r_{\alpha}s^{-\alpha}] }{s^2[s+r+r_{\alpha}s^{1-\alpha}]} \\
&&+\frac{2(\langle v(0)^2\rangle- v_T^2 )}{s^2[s+r+r_{\alpha}s^{1-\alpha}]} 
 \Big [\alpha r_{\alpha}s^{-\alpha} \nonumber \\
&&+\frac{[1+(1-\alpha)r_{\alpha}s^{-\alpha}][s+r_{\alpha}s^{1-\alpha}] }{s+r+r_{\alpha}s^{1-\alpha}}
 \Big ]\nonumber
\;.
\end{eqnarray}
From this result, it becomes immediately clear that for the equilibrium initial
preparation with $\langle v^2(0)\rangle =v_T^2$, diffusion is described by the twice
integrated $K^{\rm (st)}_v(t)$, as it was already established above.
Hence, the self-consistency test is successfully passed.
However, 
for a nonequilibrium initial preparation, the result is different.
Remarkably, it has also a different asymptotics. Namely,
\begin{eqnarray}\label{asympt}
&&\langle x^2(t)\rangle 
\sim 2\tau_r^2\frac{v_T^2+\alpha(\langle v^2(0)\rangle-v_T^2)}{\Gamma(2+\alpha)}(t/\tau_r)^{1+\alpha}\\
&& +2 \tau_r^2(1-\alpha)(\langle v^2(0)\rangle-v_T^2)(t/\tau_r)^{2\alpha}/[\Gamma(1+2\alpha)]\nonumber
\end{eqnarray}
where we kept only the leading and sub-leading terms in the limit $t\to\infty$.
The first term in (\ref{asympt}) dominates for any $0<\alpha<1$
and it displays an asymptotical dependence on the initial conditions. Such a
dependence 
is clearly a non-ergodic feature. Remarkably, for $\langle v^2(0)\rangle=0$, we obtain
the same asymptotical renormalization factor $1-\alpha$, as one derived above from
the kinetic equation (\ref{FBGE}). We recall once again that within (\ref{FBGE}), to consider 
a truly initially equilibrium velocity
preparation is simply impossible, even if to take $f(x,v,0)=\delta(x)f_M(v)$. 

\section{Discussion, Summary and Conclusions}

In this paper, we introduced two fractional generalizations of Bhatnagar, Gross, and Krook
kinetic equation in the phase space based on the picture of scattering process having finite
mean time intervals between scattering events, however, a divergent variance.
These novel fractional kinetic equations correspond to a L\`{e}vy walk in the velocity space
characterized by simplest fractional relaxation equations for the velocity variable
possible under the stated requirement of finite $\langle \tau\rangle$. In other words, they
provide a fundamental fractional kinetic 
model of general interest and applicability. The first 
fractional kinetic equation (\ref{FBGE}) is closely related to the kinetic equation
by Friedrich \textit{et al.} by taking the retardation effects into account. The
form of the scattering term is, however, very different. We have it in the form first 
suggested by the Bohm and Gross, while Friedrich \textit{et al.} have the limiting form
of the KFP equation. Moreover, we have a finite mean residence time between
scattering events whose inverse defines a mean scattering rate $r$. 
This is the second profound
difference.  The solution of (\ref{FBGE}) 
reproduces, however, asymptotically the result by Friedrich \textit{et al.} in the formal limit 
$\langle \tau\rangle \to \infty$, obtained earlier for a very different scattering model. 
This is a very interesting and important feature. It allows to clarify mathematically
rigorously if it is possible at all, in principle, neglect retardation effects as it was
done implicitly in the fractional KFP equation by Barkai and Silbey.
The correct result has for $\langle \tau\rangle =\infty$ the ballistic asymptotics
$\langle x^2(t)\rangle \sim v_T^2(1-\alpha)t^2$, the same as 
in \cite{FriedrichPRL06,FriedrichPRE06}, while
the neglection of the retardation effects in our fractional BGK equation leads
to the same incorrect result featuring fractional KFP equation without
retardation. This incorrect result is very perplexing and misleading indeed 
because it is looking
like one obtained by the double integration of the velocity autocorrelation
function 
$K_v(t,0)= K_v(t+t_{\rm ag},t_{\rm ag})|_{t_{\rm ag}=0}=v_T^2E_{\alpha}(-r_{\alpha}t^{\alpha})$, 
obtained at the zero value of the time-age variable, $t_{\rm ag}=0$. 
The treatment in the velocity
space allowed to locate and fix the problem. Namely, in the limit  $\langle \tau\rangle =\infty$, 
the correct stationary
autocorrelation function of velocities is just a constant, whose twice integration yields
the ballistic diffusion $\langle x^2(t)\rangle \sim v_T^2 t^2$. Hence, the
original version of the fractional Kramers-Fokker-Planck equation, 
which neglects the retardation effects, 
has indeed a profound defect, as it was  already revealed and
corrected by Friedrich \textit{et al.} \cite{FriedrichPRL06,FriedrichPRE06}. 
Furthermore, the different 
from $1-\alpha$ prefactor reflects the very fact that our kinetic equation (\ref{FBGE}), as well as the
fractional KFP equation by Friedrich \textit{et al.} both correspond to a very
nonstationary setup, where the time-evolution of the particles distribution function
starts from the scattering events experienced by all the particles 
at the same time $t_{\rm ag}=t_0=0$. 
Arguably, such an initial preparation is difficult, if 
possible in principle, to realize experimentally. Fortunately, in the case of finite 
$\langle \tau\rangle$, and, especially, in the important limit $\langle \tau\rangle\to 0$, i.e.
in  the limit of infinite mean scattering rate $r\to\infty$, a quasi-stationary and fully
aged description is possible with $t_{\rm ag}\to\infty$. Here, the scattering events
started in the infinite past, i.e. the system was pre-equilibrated, even though initially
it can be still very far from the equilibrium, with the initial velocity distribution
$p(v,0)$ very different from the Maxwellian distribution $f_M(v)$ finally enforced
by the particles of the thermal bath. Our second fractional BGK equation (\ref{FBGE2})
does correspond to such an initial pre-equilibration. We confirmed this by using it
to find the force-free $\langle x^2(t) \rangle$, which indeed corresponds to 
$K_v^{\rm (st)}(t)$ found in the velocity subspace from the underlying L\`{e}vy walk, provided
that the initial distribution of velocities is Maxwellian. Interestingly,
both retarded and non-retarded versions of the fractional kinetic equation
is the phase space do correspond to one and the same fractional kinetic equation
in the velocity subspace. This is a reason why the treatment in the whole phase
space is so important. Interestingly, using out-of-equilibrium initial $p(v,0)$
in (\ref{FBGE2}) does modify the asymptotic behavior of diffusion. It becomes
different from one following from $K_v^{\rm (st)}(t)$ by a factor, which, interestingly
enough, for $\langle v^2(0)\rangle=0$ becomes $1-\alpha$, i.e. the same which follows
from (\ref{FBGE}). Also very important is that the neglect of the retardation effects
in both equations (\ref{FBGE}) and (\ref{FBGE2}) leads to a completely wrong
result, which can be obtained by twice integrating $K_v(t,0)$. Instead of 
asymptotic superdiffusion $\langle x^2(t) \rangle\propto t^{1+\alpha}$, one
obtains just the normal diffusion $\langle x^2(t) \rangle\propto t$, which
misleadingly implies that in order to have asymptotic superdiffusion the
condition $\langle \tau\rangle=\infty$ is indispensable. This is, of course, completely
wrong. As a matter of fact, the neglect of retardation effect results in
the very same subtle defect which features the fractional KFP equation by Barkai
and Silbey. Strikingly enough, this defect persists even in the 
$\langle \tau\rangle\to 0$ limit. Hence, the retardation effects can
never be neglected in fractional kinetics. Another very interesting feature is that
in the limit $\langle \tau\rangle\to 0$,  
$K_v^{\rm (st)}(t)=v_T^2E_{1-\alpha}[-(t/\tau_r)^{1-\alpha}]$, which reminds the
result by Barkai and Silbey for $K_v(t,0)$ in the limit 
$\langle \tau\rangle\to \infty$. The differences are, however, profound. First, $1-\alpha$
instead of $\alpha$, and a very different relaxation scale $\tau_r$.

The proposed fractional kinetic equations are aimed for use in the externals 
force fields $F\neq 0$.
Here, the further comments are required. First, in this case one should, strictly speaking,
also take into account the retardation in the velocity variable, i.e. instead
of e.g. $f(x,v,t')$ we will have $f(X(t,t'),V(t,t'),t')$ in the scattering
term written in the form with a singular memory kernel
(without use fractional substantial derivative). Here,
$X(t,t')=x-\int_{t'}^t v(\tau)d\tau$, and 
$V(t,t')=v-\int_{t'}^t F(x(\tau),\tau)d\tau/M$. Hence, $X(t,t')=x-v(t-t')$ and 
$V(t,t')=v-F(t-t')/M$ used by Friedrich \textit{et al.} \cite{FriedrichPRL06} 
in this case is only an approximation,
which physically is rather questionable in the limit $\langle \tau\rangle\to\infty$.
Second, we suppose that for sufficiently small $\langle \tau\rangle$ and a large mass $M$,
we can yet totally neglect the retardation in the velocity variable, and approximate
$X(t,t')\approx x-v(t-t')$. The validity of this approximation should
be further tested on practical examples.
With this warning and reservation, the readers are invited to follow
the described research pathway and to use the novel kinetic equations in their
own research work. The case of the corresponding fractional dynamics driven by external
force fields is expected to bring about further insights and surprises.

\section*{Acknowledgment} 
Funding of this research by the Deutsche Forschungsgemeinschaft, Grant
GO 2052/3-1 is gratefully acknowledged.


\end{document}